%% file: deshpande_barmish_final.tex
\documentclass[letterpaper, 10 pt, conference]{ieeeconf}
\usepackage{amsmath}
\usepackage{amssymb}
\usepackage{amsfonts}
\usepackage{color}
\usepackage[sort,compress]{cite}
\usepackage{chngcntr}
\usepackage[normalem]{ulem}

\counterwithin{paragraph}{section}

\usepackage{amsthm}
\usepackage{etoolbox}
\usepackage[neverdecrease]{paralist}
\usepackage{graphicx}
\usepackage{myStyle}
\IEEEoverridecommandlockouts                              
\begin{document}
\title{\bf A Generalization of the Robust Positive Expectation\\ Theorem for Stock Trading via Feedback Control
} 
\author{\large Atul Deshpande$^{1}$ and B. Ross Barmish$^{2}$
\thanks{\hspace{-.25cm}${}^1$\hspace{1pt}Atul Deshpande is a graduate student working towards his doctoral dissertation in the Department of Electrical and Computer Engineering, University of Wisconsin, Madison, WI 53706.
        {\tt\small atul.deshpande@wisc.edu}}%
\thanks{\hspace{-.25cm}${}^2$\hspace{1pt}B. Ross Barmish is a faculty member in the Department of Electrical and Computer Engineering, University of Wisconsin, Madison, WI 53706.
        {\tt\small bob.barmish@wisc.edu}}%
        }
\maketitle
{
\begin{abstract}                
 The starting point of this paper is the so-called \emph{Robust Positive Expectation} (RPE) Theorem, a result which appears in literature in the context of Simultaneous Long-Short stock trading. 
 This theorem states that using a combination of two specially-constructed linear feedback trading controllers, one long and one short, the expected value of the resulting gain-loss function is guaranteed to be robustly positive with respect to a large class of stochastic processes for the stock price.
The main result of this paper is a generalization of this theorem.
Whereas previous work applies to a single stock, in this paper, we consider a pair of stocks.
To this end, we make two assumptions on their expected returns. The first assumption involves price correlation between the two stocks and the second involves a bounded non-zero momentum condition.
With known uncertainty bounds on the parameters associated with these assumptions, our new version of the RPE Theorem provides necessary and sufficient conditions on the positive {feedback parameter}~$K$ of the controller under which robust positive expectation is assured. 
We also demonstrate that our result generalizes the one existing for the single-stock case. Finally, it is noted that our results also can be interpreted in the context of pairs~trading.
\end{abstract}
}
%
%
\section{Introduction}
The primary motivation for this paper is the so-called \emph{Robust Positive Expectation} Theorem for Simultaneous Long-Short (SLS) trading of a single stock; see~\cite{SLS2} and~\cite{SLS8}. This result is a stochastic  version of an arbitrage theorem originally introduced for continuously differentiable stock prices in~\cite{SLS1}. 
It tells us that a combination of two controllers, one for the long trade and one for the short trade, provides a guarantee that the expected value of the gain-loss function is robustly positive with respect to a family of underlying stock prices which are Geometric Brownian Motions (GBM) with unknown drift~$\mu$ and unknown volatility~$\sigma$. Whereas robust portfolio balancing strategies have been presented in papers such as~\cite{cover}, the earliest contribution we find on robust positive expectation can be found in papers such as~\cite{RPE} and other related work by the same authors, {such as~\cite{DokBook}}. In contrast to the above, we focus here on the linear feedback control framework which is covered in papers such as~\cite{SLS1,SLS2,SLS8} and~{\cite{SLS3,SLS4,SLS5, SLS6, SLS7, BauNew}}.
\begin{sloppypar}
The body of literature motivating this paper includes a number of flavors for the underlying stock prices and the control structure. For example, in reference~\cite{SLS5}, robustness results are given for stock prices generated by Merton's jump diffusion model and references~{\cite{SLS7, SLS6, BauNew}} address variants of the SLS controller for the discrete-time case. 
To conclude this brief survey, we note that most of the literature cited above falls within the robust control paradigm formulated in~\cite{IFAC}. 
Less closely related to this line of research are references~\cite{{1,11,13,14,Pairs2,PairsQ,Pairs3}}, which, unlike the papers on robust control, are based on rather specific stock-price~models. {For example, in~\cite{1}, stock prices are modeled as GBM processes coupled by a finite-state Markov chain, and in~\cite{14}, trading signals are modeled as Ito processes based on GBM models. On the other hand, in~\cite{11, 13, Pairs2, PairsQ, Pairs3}, either the asset being traded or a relationship between multiple assets, is modeled as a mean-reverting Ornstein-Uhlenbeck process.}
\end{sloppypar}
Whereas the SLS literature focuses on trading shares of a single stock,
in this paper, we consider scenarios involving simultaneously trading two stocks.
{One simple method to extend the single-stock theory to two stocks would be to implement separate SLS controllers for each stock. That is, a robustly positive expected (RPE) gain for each stock individually implies that the pairs trade has RPE too.
In this paper, we study a different approach for trading a pair, where one arm of a controller goes long on one of the stocks and the other arm goes short on the other stock.}
{This new control structure is motivated by the desire to exploit correlated price behavior between two stocks rather than treating them separately.}
{
To this end, we make certain assumptions on the stock dynamics, namely the satisfaction of}
{\it directional correlation} and {\it bounded momentum} conditions. 
Letting~$g(N)$ denote the cumulative gain or loss up to stage~$N$, we describe a generalized SLS controller with {feedback parameter}~$K$, which is constructed using the known uncertainty bounds.
Our main result for the two-stock case provides necessary and sufficient conditions on~$K$ under which robust satisfaction of the~condition~$
\mathbb{E}[g(N)]>0
$ 
is guaranteed with respect to parameter variations associated with the conditions above.
We also show how these results generalize the RPE Theorem for the single-stock scenario.\\[5pt]
Given that our formulation is aimed at two stocks with correlated price dynamics, this paper provides a new perspective on ``pairs-trading'' literature. Unlike this literature, however, we do not include assumptions of price~reversion, either through reliance on models such as those of Ornstein-Uhlenbeck as seen in~\cite{Pairs3,Pairs2,13,PairsQ} or more general models for the spread function as in~\cite{yamada} and~\cite{Pairs1}.
\subsubsection*{Existing Result Being Generalized}
\begin{figure}[!t]
\centering
\includegraphics [width=3.4in]{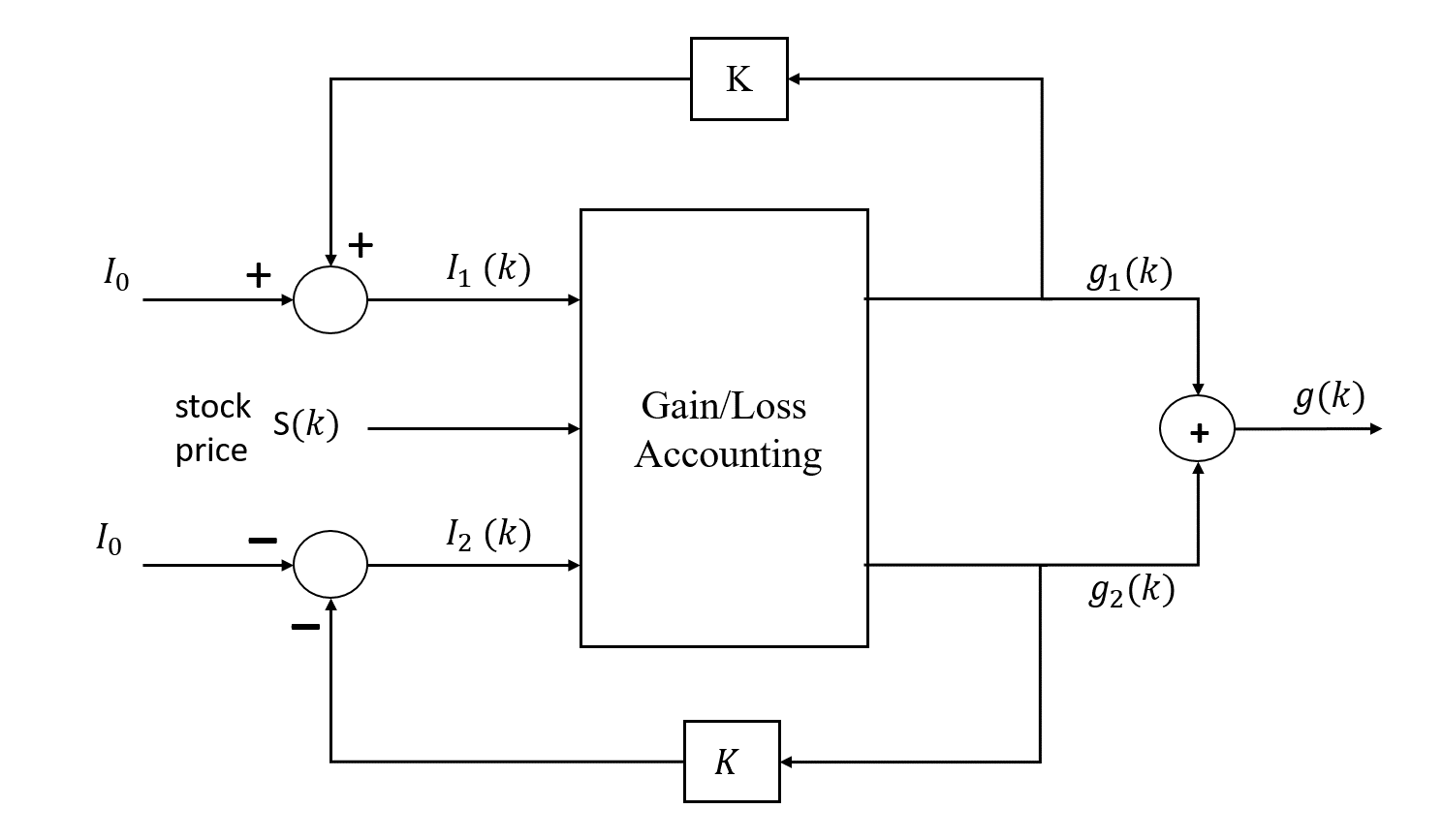}
\caption{The SLS Controller}
\label{prior_block}
\vspace{-1.5em}\end{figure}

The take-off point for this paper is the Robust Positive Expectation Theorem for an SLS controller used to trade a single stock. Indeed, assuming a stock with prices represented by a discrete-time stochastic price process~$S\left(k\right)$ over~${k=0,1,\dots N}$, let~$\rho\left(k\right)$ denote the return in the~$k$-th period; i.e.,
{\small\begin{equation*}
\rho\left(k\right) \doteq \frac{S\left(k+1\right) - S\left(k\right)}{S\left(k\right)},
\end{equation*}}\\[-10pt]
are taken to be independent, with an unknown constant~mean~$
\mu\doteq\mathbb{E}[\rho\left(k\right)]
$
for $k=0,1,2,\dots,N-1$.\\[5pt]
Given the setup above, the Simultaneous Long-Short (SLS) controller, depicted in Figure~\ref{prior_block}, determines the net investment level~$I\left(k\right)$ in the stock at stage~$k$. This is accomplished by summing the outputs of two linear time-invariant controllers. The first uses an initially positive~$I_1\left(k\right)$ for the long trade and the second uses an initially negative~$I_2\left(k\right)$ for the short trade. To elaborate, a \emph{long} position~$I_1(k)>0$ represents the trader holding the appropriate number of shares of the stock and making profit as~$S\left(k\right)$ increases. On the other hand, a \emph{short} position~$I_2(k)<0$ leads to a profit when there is a decrease in the stock price.
We take
{\small\begin{align*}
 I_1\left(k\right) \doteq I_0 + K g_1\left(k\right);\;\;
 I_2\left(k\right) \doteq -I_0 - K g_2\left(k\right)
\end{align*}}\\[-10pt]
with initial investment~${I_0>0}$, {feedback parameter}~$K>0$ and~$g_1\left(k\right)$,~$g_2\left(k\right)$ being the cumulative gain-loss functions of the two controllers, with initial values~$g_1(0)=g_2(0)=0$. Subsequently, the trader's net investment level in the stock~$I\left(k\right)$ is obtained~as
{\small
$$
I\left(k\right) = I_1\left(k\right) + I_2\left(k\right) = K\left(g_1\left(k\right) - g_2\left(k\right)\right).
$$}\\[-10pt]
\\[-10pt]
The robust positive expectation result from which we take off tells us: Except for the {degenerate} break-even case obtained with~$\mu=0$, the cumulative gain-loss function 
{\small
$$
g\left(k\right) = g_1\left(k\right) + g_2\left(k\right)
$$}\\[-10pt]
is robustly positive in expectation. That is, without knowledge of~$\mu$, the condition~$
\mathbb{E}[g\left(N\right)]>0$ is satisfied.
Furthermore, as seen in existing work such as~\cite{SLS7}, the expected gain-loss function is explicitly given~by
{\small
$$
\mathbb{E}[g\left(N\right)] = \frac{I_{0}}{K}\left[ \left(1 + K \mu \right)^N + \left(1 - K \mu \right)^N - 2\right],
$$}\\[-10pt]
with the positivity of the above expression guaranteed for all non-zero~$\mu$ by virtue of the basic fact that~$\left(1+x\right)^N +\left(1-x\right)^N>2$ for all~$x\neq 0$ and~$N\geq 2$. Since this result is the starting point for our current work, for the sake of a self-contained exposition, we provide an elementary derivation of the formula for~$\mathbb{E}[g(N)]$ above in the~appendix.
\section{Two-Stock Setup and Market Assumptions}
In this section, we consider two stocks instead of one and now describe the assumptions which are in force. These assumptions are not only on the price processes for the two stocks, but also on the market within which we operate.
\subsubsection*{Stock Price Dynamics}
We consider stocks~$S_1$ and~$S_2$ with stochastically varying prices~$S_1\left(k\right)$ and~$S_2\left(k\right)$ respectively for~$k=0,1,2,\dots, N$ and~$N>1$.
The returns on the stocks, given by
\begin{equation*}
\rho_i\left(k\right) \doteq \frac{S_i\left(k+1\right) - S_i\left(k\right)}{S_i\left(k\right)}
\end{equation*}
for~$i=1,2$ are respectively assumed to be independent for~$k=0,1,2, \dots, N-1$, with constant~means~$
{\mu_1 \doteq \mathbb{E} [\rho_1\left(k\right)]},
{\mu_2 \doteq \mathbb{E} [\rho_2\left(k\right)]}.
$
The relationship between these returns are assumed to satisfy the following~conditions:
\subsubsection*{Directionally Correlated Returns Assumption} We assume that there exists a constant~$\beta\neq 0$ such that~$
{\mu_2 = \beta {\mu_1}}
$~with~
{\small
$$
\beta = \left(1+\varepsilon\right) \beta_0,
$$}\\[-10pt]
$\beta_0\neq 0$ known to the trader and~$\varepsilon$ uncertain, with known~bounds~$
0\leq\varepsilon\leq\varepsilon_{max}.
$
Note that the above implies that~${\text{sign}\;\beta_0=\text{sign}\;\beta}$ for all admissible~$\varepsilon$, that is, there is uncertainty in the magnitude of~$\beta$, but not its sign. 
\subsubsection*{Bounded Non-Zero Momentum Assumption} It is assumed that
there are positive constants~$\mu_{min}$ and~$\mu_{max}$ known to the trader, such that
$$
{\mu_{min}\leq|\mu_1|\leq \mu_{max}}.
$$

{\bf Remarks:} The bounds on~$\mu_1$ above in combination with the assumption of directionally correlated returns lead to bounds on~$\mu_2$ given by
$${0<|\beta_0| \mu_{min}\leq|\mu_2|\leq (1+\varepsilon_{max})|\beta_0| \mu_{max}}.$$
For the special case when the two stocks are one and the same,~$\beta\equiv 1$ {with $\varepsilon_{max}=0$}, and this formulation reduces to a restricted version of the single-stock problem {described in~literature}.
\subsubsection*{Idealized Market Assumptions}
The trading is assumed to be carried out under idealized market conditions. 
That is, there are no transaction costs such as brokerage commission, fees or taxes for buying or selling shares.
For such a market, it is also assumed that there is perfect liquidity; i.e., there is no gap between the bid and ask prices, and the trader can buy or sell any number, including \emph{fractions}, of shares as desired at the currently traded price. That is, the trader is a price taker, investing small enough amounts so as not to affect the prices of the stocks.
These assumptions are similar to those made in finance literature in the context of  ``frictionless markets'' going as far back~as~\cite{Merton}.
\subsubsection*{Leverage, Margin and Interest}
In practice, brokers usually impose a limit on the investment levels based on the account value~$V(k)$. For example, a trader may be bound by the constraint~$
{|I_1\left(k\right)|+|I_2\left(k\right)| \leq \gamma V\left(k\right)},
$
where~$\gamma\geq 1$ denotes the so-called leverage which is extended. In the theory to follow, it is assumed that leverage is never a limiting factor. That is, sufficient resources are available to cover any desired investment levels in the respective stocks. Accordingly, issues involving margin interest are not in play. Finally, it is noted that there is no mention of ordinary interest on idle cash in the trading account. The explanation for this is that the results in this paper focus entirely on the gains and losses~$g_1(k)$ and~$g_2(k)$ which are attributable to~trading.
\section{The Two-Stock Controller}
Beginning with stocks~$S_1$ and~$S_2$, the two-stock generalized SLS controller which we now describe has the same structure as the one in Figure~\ref{prior_block}. However, for this more general case, we have both stock prices as inputs to the controller, allow for different initial investments~$I_{0,1}, I_{0,2}$ instead of~$I_0$ and have~$K_1, K_2$ instead of~$K$. The linear feedback controllers investing~$I_1\left(k\right)$ in~$S_1$ and~$I_2\left(k\right)$ in~$S_2$ are given by
{\small\begin{align*}
  I_1\left(k\right) &\doteq I_{0,1} + K_1 g_1\left(k\right);\\
  I_2\left(k\right) &\doteq -I_{0,2} - K_2 g_2\left(k\right),
 \end{align*}}\\[-10pt]
with parameters~$I_{0,i}$ and~$K_i$ chosen by the trader as explained below and with~$g_1\left(k\right)$ and~$g_2\left(k\right)$ being the cumulative gain-loss functions of the investments with initial values of~${g_1\left(0\right)=g_2\left(0\right)=0}$.
\subsubsection*{Choice of Parameters}
We first select initial investment parameter~$I_0>0$ and {feedback parameter}~$K>0$. Then, the two controllers, defined in terms of these two parameters, have initial investment levels
$$
I_{0,1} \doteq I_0;\;\;I_{0,2} \doteq \frac{I_0}{\beta_0}
$$
and {feedback parameter}s
$$
K_{1} \doteq K;\;\;
K_{2} \doteq \frac{K}{\beta_0}.
$$

{\bf Remarks:} We observe that the choices of~$I_{0,i}$ and~$K_i$ above compensate against the differing momenta of the two~stocks. When~${\beta_0>0}$, notice that the initial investments satisfy~${I_1\left(0\right)>0}$ and~${I_2\left(0\right)<0}$. However, the signs of one or both these quantities may change at a later stage~$k$. 
 Thus, despite being initially long on~$S_1$ and short on~$S_2$, our stock positions at later stage~$k$ can be different. 
 A similar statement can be~made for~$\beta_0<0$.
 \subsubsection*{Starting Point for the Analysis} A simple adaptation of the single-stock formula in Section I leads us to
{\small\begin{align*}
 \mathbb{E}[g\left(N\right)] &= \frac{I_{0,1}}{K_1}\left[\left(1 + K_1\mu_1 \right)^N -1\right] + \frac{I_{0,2}}{K_2}\left[\left(1 - K_2\mu_2 \right)^N -1\right] \\[3pt]
 &= \frac{I_0}{K}\left[\left(1 + K\mu_1 \right)^N + \left(1 - K\mu_1\left(1+\varepsilon\right) \right)^N -2\right]\\[3pt]
 &\doteq {G_{_N}}(K,\mu_1,\varepsilon)
\end{align*}}\\[-10pt]
for the two-stock case.
Note that with~$\varepsilon_{max}=0$, the formula above reduces to the one for the single-stock case. The notation~${G_{_N}}(K,\mu_1,\varepsilon)$ above making the dependence on~$K$,~$\mu_1$ and~$\varepsilon$ explicit will be useful in the sequel for presentation and proof of the results.
\section{Main Results}
In the theorem to follow, we characterize the set of~$K$ leading to the satisfaction of the robust positive expectation of~$g(N)$ with respect to~$\mu_1$ and~$\varepsilon$ within their respective bounding sets.
We also provide a corollary which leads to the recovery of the existing single-stock result when~$\beta_0=1$ and~$\varepsilon_{max}\rightarrow 0$. All the proofs for the results in this section are furnished in~Section~V.
\subsubsection*{Robust~Positive~Expectation~Theorem} \emph{Suppose two stocks~$S_1$ and~$S_2$ have directionally correlated returns and satisfy the bounded non-zero momentum condition, with associated uncertainty bounds~${0\leq \varepsilon\leq \varepsilon_{max}}$ and~${0<\mu_{min}\leq |\mu_1| \leq \mu_{max}}$.
Then, for~$N$ odd, the two-stock generalized SLS controller with~${K>0}$ guarantees robust satisfaction of the condition~$G_{_N}\left(K,\mu_1,\varepsilon\right)> 0$
 for all admissible~$\mu_1$ and~$\varepsilon$, if and only if$$
  {G_{_N}}(K,\mu_{min},\varepsilon_{max})>0;\;\;{G_{_N}}(K,\mu_{max},\varepsilon_{max})>0.$$
  For~$N$ even, robust satisfaction is guaranteed if and only if~either~${
  K>{2^\frac{1}{N}-1}/{\mu_{min}}},
 $ or when both~{\small ${
  K\leq {1}/({\mu_{min}(1+\varepsilon_{max})})}$ and~$${{G_{_N}}(K,\mu_{min},\varepsilon_{max})>0}.$$
}}\\[-5pt]
{\bf Remarks:}
\begin{samepage}
To accurately estimate the set of~$K$ which guarantees satisfaction of the robust positive expectation conditions above, as demonstrated in Section VI, we can simply conduct a parameter sweep over a suitably large range with~$K>0$.
Unlike existing results for the single-stock case, one possible outcome is that the set of~$K$ satisfying the theorem requirements is empty. 
This can occur when the uncertainty bounds are ``too large.'' 
The corollary below is apropos to the special case when both stocks are one and the same; i.e.~$\beta_0=1$ and we consider~$\varepsilon_{max}\rightarrow 0$ to recover the existing result for the single-stock case is presented below.
\end{samepage}
\subsubsection*{Corollary}\emph{
Given any~$K>0$, for~$\varepsilon_{max}$ suitably small, robust satisfaction of the condition~$\mathbb{E}[g\left(N\right)]>0$ for all admissible~$\mu_1$ and~$\varepsilon$ is~guaranteed.}
\section{Proof of the Theorems}
\input{epsc_proof}
\section{Illustrative Example}
This section demonstrates the construction and use of the two-stock controller for a toy example with Geometric Brownian Motion (GBM) as the underlying price process.
For the first stock, the discrete-time GBM which we use for daily updates is described by
$$
\frac{S_1(k+1) - S_1(k)}{S_1(k)} = \mu_1 + \sigma_1 w_1(k)
$$
where~$\mu_1$ is the drift, $\sigma_1$ the volatility and the $w_1(k)$ are independent standard normal random variables. To illustrate the application of the Robust Positive Expectation Theorem, we begin with uncertainty bounds~$\mu_{min} = 0.00055$ and~$\mu_{max} = 0.002$. Assuming each time step above represents a daily return, these bounds correspond to variations of $25\%$ and $65\%$ respectively on an annualized basis. In addition, the two stocks are assumed to be directionally correlated with a nominal $\beta_0=1$ and uncertainty bound~$\varepsilon_{max}=0.8$;~i.e.,~$1\leq \beta \leq 1.8$. Therefore, the discrete-time GBM for the second stock is described by uncertain drift~$\mu_2 = (1+\varepsilon)\mu_1$ and volatility~$\sigma_2$ as
{\small\begin{align*}
 S_2\left(k+1\right) = \left(1+(1+\varepsilon)\mu_1 + \sigma_2 w_2\left(k\right)\right)S_2\left(k\right)
\end{align*}}\\[-10pt]
with the~$w_2(k)$ being independent random variables, each having standard normal distribution.
It is important to note here that the trader does not know the direction of the underlying stock-price movement. 
That is, the sign of~$\mu_1$ is unknown; only the bounds on~$|\mu_1|$ are available. In the analysis to follow, we take $N=125$, which represents, given the daily update equations, about six months of trading.
\subsubsection*{Controller Design}
 Beginning with initial investment $I_0=10,000$ in dollars, we seek to find a suitable~$K>0$ satisfying the requirements of the RPE Theorem. Since $N$ is odd, we work with the inequalities
 $$
 {G_{_N}}(K,\mu_{min},\varepsilon_{max})>0;\;\;{G_{_N}}(K,\mu_{max},\varepsilon_{max})>0.
 $$
 For the given uncertainty bounds, we obtain the~conditions
 {\small\begin{align*}
  (1+0.00055K)^{125} + (1-0.00099K)^{125}&>2;\\[3pt]
  (1+0.002K)^{125} + (1-0.0036K)^{125}&>2
 \end{align*}}\\[-10pt]
 as being necessary and sufficient for robust positive expectation.
 Conducting a parameter sweep with~$K>0$, the inequalities above are satisfied if and only if~$
 6.33<K<1250.$
 \subsubsection*{Some Practical Considerations}
 In this subsection, we continue the analysis of the example above by introducing some practical considerations into a simulation not covered by the theorem. Recalling the discussion of leverage in Section II, we now take~$\gamma=2$, use~$K=25$ and assume an initial account value~$V(0)=10,000$ in dollars.
Then, to remain in compliance with the leverage constraint, any time the controller in Section III encounters~$|I_1\left(k\right)|+|I_2\left(k\right)|> \gamma V\left(k\right)$, the two investments are scaled back using the formula
{\small\begin{align*}
 I_i\left(k\right)&=\frac{I_i\left(k\right)}{|I_1\left(k\right)|+|I_2\left(k\right)|}\gamma V\left(k\right);\;\; i=1,2.
\end{align*}}\\[-20pt]
\subsubsection*{Controller Performance Over a Sample Path}
\begin{figure}[!t]
\centering
\includegraphics [width=3.4in]{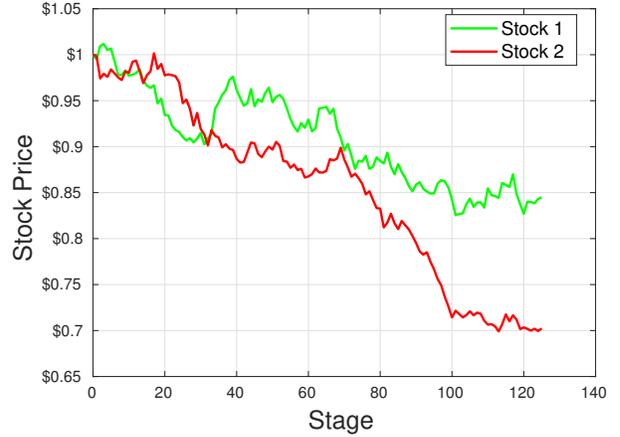}
\caption{Simulated Prices Along One Sample Path}
\label{prices1}
\vspace{-1.0em}\end{figure}

\begin{figure}[!t]
\centering
\includegraphics [width=3.4in]{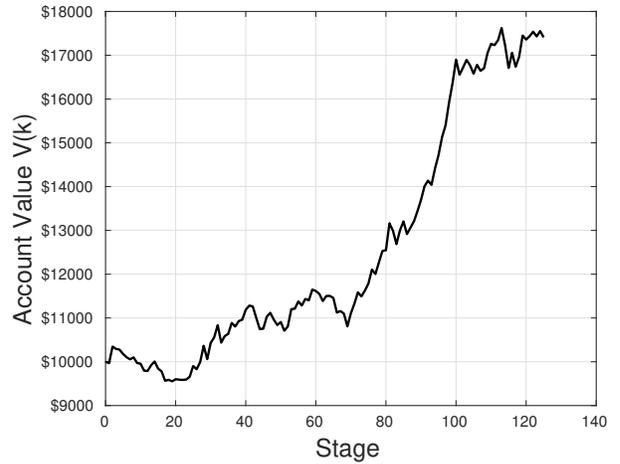}
\caption{Controller Gain-Loss Function for Scenario in Fig.~\ref{prices1}}
\label{k0p5}
\vspace{-1.5em}\end{figure}

For our simulations, we use daily volatilities of~$\sigma_1=\sigma_2=.0094$, and admissible GBM drift parameter~$\mu_1=-0.0017$, with~$\varepsilon=0.6$. This corresponds to~$\beta = 1.6$ between the two stocks and a drift of~$\mu_2=0.0027$.
{First, we illustrate} the controller performance for a single sample path  for each of the two stock prices; see Figure~\ref{prices1} where we observe price declines of approximately~$17.5\%$ for $S_1$ and~$25\%$ for $S_2$ over the trading period. 
Figure \ref{k0p5} shows the performance of the controller. We see an overall return of~$74\%$ on the initial~$\$10,000$ during the trading period. It is also noteworthy that most of the gains come in the trading period between stages~70 and~100. This is the period when the largest downward stock price movement~occurs.\\[-10pt]
\subsubsection*{Aggregate~Statistics over Many Sample Paths}
\begin{figure}[!t]
\centering
\includegraphics [width=3.4in]{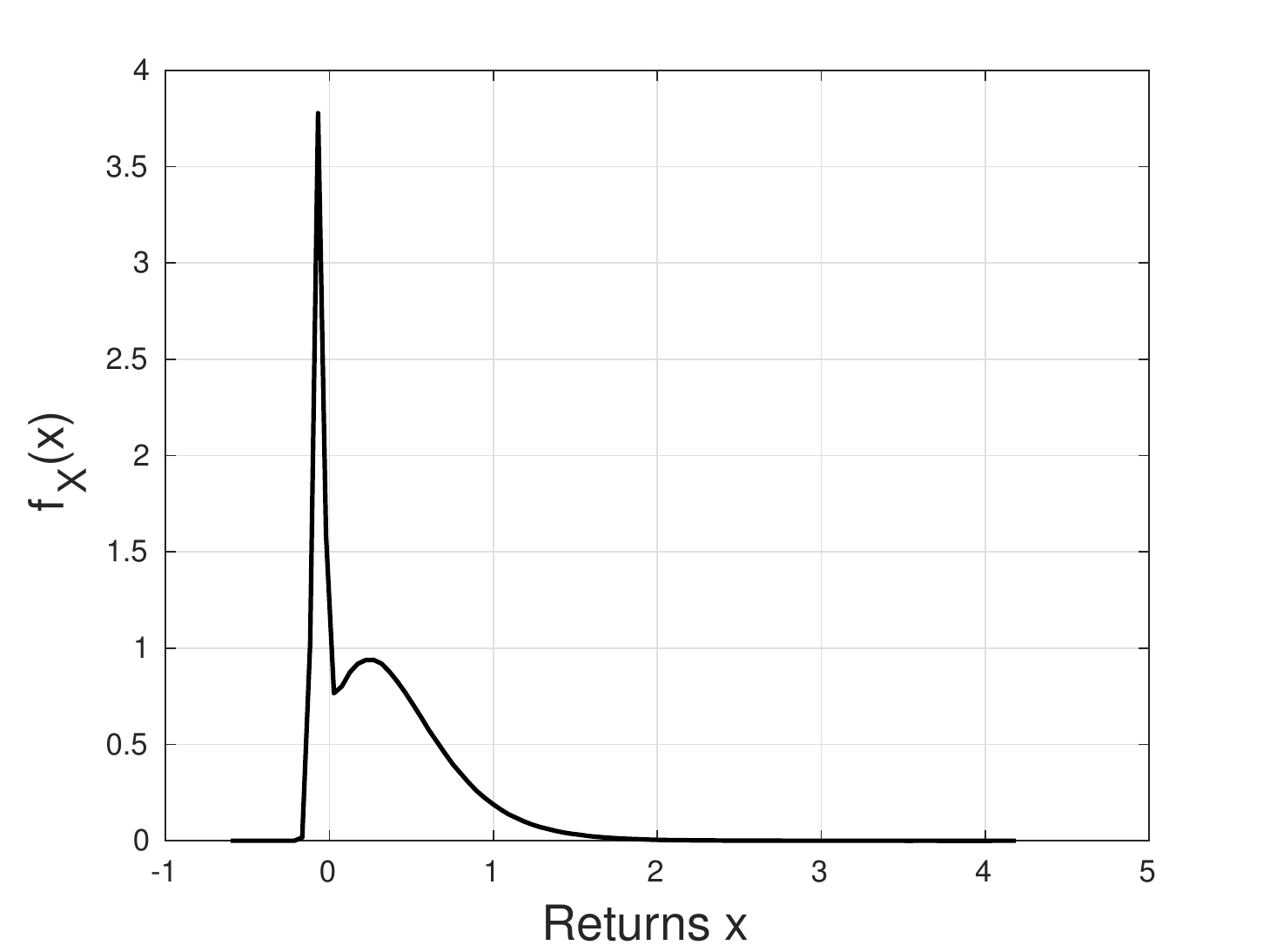}
\caption{Probability Distribution of Returns}
\label{hist}
\vspace{-1.5em}\end{figure}

Now, instead of the single sample path analysis, we consider the performance against the entire family of GBM processes under consideration. We now calculate the returns~$
X={V(N)-V(0)}/{V\left(0\right)}
$ 
using one million sample paths with~$\mu_1$ and~$\varepsilon$ chosen using the uniform distribution over their respective admissible ranges.
Figure~\ref{hist} shows the empirically estimated probability density function of~$X$.
The controller yields an average return of about~$30.2\%$ and a median return of about~$22.7\%$ with a probability of profit of~$0.69$. Interestingly, the statistics indicate positive expected value for~$g(N)$ even with the added leverage~constraints.
Most notably, even among the unprofitable scenarios, we observe that the controller limits the losses. For example,~99.99\% of all the unprofitable sample paths show losses limited to less than 10\% of the initial account value. 
\section{Conclusion}
The main result in this paper is a new version of the Robust Positive Expectation (RPE) Theorem for the case of trading two directionally correlated stocks with bounded non-zero momenta.
Given the uncertainty bounds~$\mu_{min}$,~$\mu_{max}$ and~$\varepsilon_{max}$, the theorem provided necessary and sufficient conditions on~$K$ under which robustly positive expected trading gain~$\mathbb{E}[g(N)]$ is guaranteed.
{If the conditions of the theorem result in no positive~$K$ satisfying the RPE condition, we deem the pair as not tradable. This reflects the fact that the uncertainty bounds are too large to enable robustness guarantees.}
\\[5pt]
By way of future research, 
a logical step would be to back-test our new two-stock controller using historical data and to compare the performance to that of traditional pairs-trading algorithms.
{It is also worth noting that the investment levels~$I_1(k)$ and~$I_2(k)$ of the two arms of our new controller evolve independently. A potential research direction involves the development of new controllers with cross-coupling in their investment levels; each controller depends on the performance of the other.}
Another interesting direction of research would be to generalize the theory presented here to a basket of more than two directionally-correlated stocks. 
\\[-10pt]
\appendix
{
Here, we obtain the expression for~$\mathbb{E}[g\left(N\right)]$ for an SLS controller operating on a single stock. Given the price process~$S\left(k\right)$ over~$k=0,1,2,3,\dots,N$, having independent returns~$\rho\left(k\right)$ with constant mean~$\mu=\mathbb{E}[\rho\left(k\right)]$,
beginning with the SLS controller~$
 I_1\left(k\right) \doteq I_{0} + K g_1\left(k\right),
 I_2\left(k\right) \doteq -I_{0} - K g_2\left(k\right)$
described in Section I with the update equations
{\small\begin{align*}
g_i\left(k+1\right) = g_i\left(k\right) + I_i\left(k\right)\rho\left(k\right)
\end{align*}}\\[-10pt]
for~$i=1,2$
and~$g_1\left(0\right)=g_2\left(0\right)=0$,
substituting for~$I_i\left(k\right)$, 
{\small\begin{align*}
g_1\left(k+1\right) = \left(1+K\rho\left(k\right)\right)g_1\left(k\right) + I_0\rho\left(k\right);\\[3pt]
 g_2\left(k+1\right) = \left(1-K\rho\left(k\right)\right)g_2\left(k\right) - I_0\rho\left(k\right).
\end{align*}}\\[-10pt]
Taking the expectation in both the equations while noting that~$\rho\left(k\right)$ and~$g_i\left(k\right)$ are independent, we~obtain
{\small\begin{align*}
\mathbb{E}[g_1\left(k+1\right)] = \left(1+K\mu\right)\mathbb{E}[g_1\left(k\right)] + I_{0}\mu;\\[0pt]
\mathbb{E}[g_2\left(k+1\right)] = \left(1-K\mu\right)\mathbb{E}[g_2\left(k\right)] - I_{0}\mu.
\end{align*}}\\[-10pt]
Since each equation above has a simple scalar state-space form~$
x\left(k+1\right) = ax\left(k\right)+bu\left(k\right)
$
with zero initial condition and constant input~$u\left(k\right)=I_0\mu$,
a straightforward calculation leads~to
{\small\begin{align*}
\mathbb{E}[g_1\left(N\right)] = \frac{I_{0}}{K}\left({\left(1+K\mu\right)^N-1}\right);\\
\mathbb{E}[g_2\left(N\right)] = \frac{I_{0}}{K}\left(\left(1-K\mu\right)^N-1\right).
\end{align*}}\\[-10pt]
Now, summing the two solutions above, we obtain 
{\begin{align*}
 \mathbb{E}[g\left(N\right)]=\frac{I_{0}}{K}\left[\left(1+K\mu\right)^N+\left(1-K\mu\right)^N-2\right].
\end{align*}}\\[-20pt]
}
 \bibliographystyle{IEEEtran}

\end{document}

%% file: epsc_proof.tex
This section can be skipped by the reader seeking to avoid technicalities. 
Recalling that~$G_{_N}(K,\mu_1,\varepsilon)$ represents~$\mathbb{E}[g(N)]$ for a fixed~$K$,~$\mu_1$ and~$\varepsilon$, the starting point for our analysis is the fact that the robust positive expectation property holds if and only if 
{\small
$$
G_{_N}(K,\mu_1,\varepsilon)>0
$$}\\[-10pt]
for all admissible pairs~$(\mu_1,\varepsilon)$. 
We first present some notation, a preliminary definition and a few lemmas which will be instrumental to the proofs to follow.
Indeed, for fixed~$\theta$, we define the polynomial
{\small
$$
 G_\theta\left(\varepsilon\right)\doteq \left(1+\theta\right)^N + \left(1-\theta\left(1+\varepsilon\right)\right)^N-2
$$}\\[-10pt]
for~${\varepsilon\geq 0}$. Note the similarity between the expressions for~$G_\theta\left(\varepsilon\right)$ and~$G_{_N}(K,\mu_1,\varepsilon)$. Indeed, when~$\theta=K\mu_1$,~$G_\theta\left(\varepsilon\right)>0$ if and only if~$G_{_N}(K,\mu_1,\varepsilon)>0$.\\[-6pt]
\subsubsection*{Definition (Critical Uncertainty Bound)}
  For a fixed~$\theta$, the \emph{critical uncertainty bound} is defined as
 {\small
$$
\varepsilon_c\left(\theta\right)\doteq \inf\{\varepsilon> 0:G_\theta\left(\varepsilon\right)\leq 0\}.
 ~$$}\\[-6pt]
 {\bf Remarks:} Given~$K$ and~$\mu_1$, the quantity~$\varepsilon_c(K\mu_1)$ tells us the smallest~$\varepsilon$ for which the expected gain~$G_{_N}(K,\mu_1,\varepsilon)$ is non-positive.
 Using the convention that the infimum over an empty set is~$+\infty$,
 if~$\theta<0$, since~$G_\theta\left(\varepsilon\right)>0$ for all~$\varepsilon>0$, we obtain~$\varepsilon_c\left(\theta\right)=+\infty$. Furthermore, when~$\theta=0$,~$G_\theta\left(\varepsilon\right)=0$ for all~$\varepsilon>0$, thus~$\varepsilon_c\left(0\right)=0$. Finally, for~$\theta>0$, notice that the continuity of~$G_\theta\left(\varepsilon\right)$ in combination with the fact that~$G_\theta\left(0\right)>0$ ensures that~$\varepsilon_c\left(\theta\right)>0$.
 The lemmas to follow more fully characterize the function~$\varepsilon_c\left(\theta\right)$ for the non-trivial case when~$\theta>0$.
 \subsubsection*{Notational Convention} In the proof to follow, there are numerous occasions where root operations are required. To avoid ambiguities attributable to non-unique or complex roots, the following notational conventions are in force: If~$X > 0$ is real and~$N$ is a positive integer, then~$X^{1/N}$ is taken to be the unique positive~$N$-th root of~$X$. For~$X < 0$ and~$N$ odd, we take~$X^{1/N} =  -|X|^{1/N}$ which is obtained using the definition for the positive variable case.  We provide no definition when~$X<0$ for~$N$ even since this case is never encountered in the sequel. There are also cases when we consider expressions of the form~$X^{1/N - m}$ for an integer~$m$. In this case, this quantity is defined as~$Y^{1/N}$, where~$Y= X^{1-mN}$, and evaluated in a manner consistent with the convention above.
 \medskip
 \begin{lemma}[Critical Uncertainty]
 Given~$\theta>0$, it follows~that
 {\small
~$$
 \varepsilon_c\left(\theta\right) = \begin{cases}
                        +\infty&N \text{ even},\; \theta>2^\frac{1}{N}-1;\\[3pt]
                        \frac{1-\left(2-\left(1+\theta\right)^N\right)^\frac{1}{N}}{\theta}-1 &\text{otherwise}.
                       \end{cases}
~$$}\\[-10pt]
\end{lemma}
{\bf Proof:} For~$N$ even and~${\theta> 2^{1/N}-1}$, since
{\small\begin{align*}
G_\theta\left(\varepsilon\right)\geq \left(1+\theta\right)^N - 2
\end{align*}
}\\[-5pt]for all~${\varepsilon>0}$, it follows that the set of~$\varepsilon$ for which~$G_\theta\left(\varepsilon\right)\leq 0$ is empty. Hence,~${\varepsilon_c\left(\theta\right) = +\infty}$.
For all other~$\theta>0$ for~$N$ odd or even,~$\varepsilon_c(\theta)$ must be the smallest finite~$\varepsilon>0$ solving the equation~$G_\theta\left(\varepsilon\right)=0$. This is easily found to be
{\small
$$
\varepsilon_c\left(\theta\right) = \frac{{1-\left(2-\left(1+\theta\right)^N\right)^\frac{1}{N}}}{\theta}-1. \qed\\[-12pt]
$$}
\subsubsection*{Definition} 
To facilitate the proof of the following lemmas, we define the function $f(\theta)$ on the set of positive~$\theta\neq 2^{1/N}-1$~as
{\small\begin{multline*}
 f\left(\theta\right) \doteq\left(2-\left(1+\theta\right)^N\right)^\frac{1}{N}-1\\
 +{\theta\left(1+\theta\right)^{N-1}\left(2-\left(1+\theta\right)^N\right)^{\frac{1}{N}-1}}.\hspace{0.5cm}
\end{multline*}
}\\[-10pt]
Furthermore, in the sequel, we also use its derivative for~$\theta\neq 2^{1/N}-1$. This is calculated to~be
{\small\begin{multline*}
f'\left(\theta\right) = 
2\left(N-1\right)\theta\left(1+\theta\right)^{N-2}\left[2-\left(1+\theta\right)^N\right]^{\frac{1}{N}-2}.
\end{multline*}}\\[-10pt]
\begin{lemma}[Monotonicity]
 For~$0<\theta<2^{1/N}-1$, the function~$\varepsilon_c\left(\theta\right)$ is monotonically increasing with derivative
 {\small
 $$
 \varepsilon'_c(\theta) = f(\theta)/\theta^2.
 $$ }
 \end{lemma}
{\bf Proof:} 
In the interval~${0<\theta< 2^{1/N}-1}$, a straightforward calculation leads to~$\varepsilon'_c\left(\theta\right)$ as given above.
To complete the proof, it suffices to show that~$f\left(\theta\right)>0$ in the interval of interest. 
Since~$f\left(0\right)=0$, and~$f'\left(\theta\right)>0$ in the interval, by inspection, it follows that~$f\left(\theta\right)>0$ for all~$\theta$ in the~interval.$\qed$\medskip
\begin{lemma}[Maximality]
 For~$N$ odd and~${\theta>2^{1/N}-1}$, the function~$\varepsilon_c\left(\theta\right)$ has a unique stationary point where it attains its maximum value,
 with its derivative
 {\small
 $$
 \varepsilon'_c(\theta) = f(\theta)/\theta^2.
 $$ }
\end{lemma}
{\bf Proof:}
For~$N$ odd and~${\theta>2^{1/N}-1}$, we show that~$\varepsilon_c\left(\theta\right)$ initially increases, thereafter achieves a maximum and decreases as $\theta$ continues to increase. 
To this end, it suffices to show that~$\varepsilon'_c\left(\theta\right)$ is initially positive, later crosses zero and thereafter stays negative. For $\theta>2^{1/N}-1$, straightforward calculation leads to~$\varepsilon'_c\left(\theta\right)$ as given above.
Indeed, it now suffices to show that its numerator~$f\left(\theta\right)$ behaves in the manner described above.
Indeed, for~$\theta>2^{1/N}-1$,~$f(\theta)$ tends to infinity as~$\theta$ approaches~$2^{1/N}-1$ from above.
However, it decreases monotonically thereafter as~$f'\left(\theta\right)$ is negative for all~$\theta>2^{1/N}-1$. 
Finally, the fact that~$f\left(\theta\right)$ eventually becomes negative is immediate since~$
\lim_{\theta\rightarrow +\infty} f\left(\theta\right) = -2.
$
Hence,~$\varepsilon'_c\left(\theta\right)$, while initially positive for~$\theta>2^{1/N}-1$, decreases to cross zero and turns negative, which in turn implies that~$\varepsilon_c\left(\theta\right)$ increases to its maximum and decreases thereafter as~$\theta$ increases to infinity. 
 This completes the~proof.~$\qed$
\subsubsection*{Proof of Robust Positive Expectation Theorem}
\input{necessity_noepsc}
To establish sufficiency, we assume feedback gain~$K>0$ satisfying the conditions in the theorem and must show that~$G_{_N}(K,\mu_1,\varepsilon)>0$ for all admissible~$\mu_1$ and~$\varepsilon$. 
To this end, we choose an arbitrary admissible pair $(\mu_1,\varepsilon)$, and divide the analysis into three~cases:\\[5pt]
\begin{samepage}
{\it Case 1 ($\mu_1<0$):} In this case, whether $N$ is even or odd, the condition~${G_{_N}(K,\mu_1,\varepsilon)>0}$ is trivially satisfied by virtue of the fact that 
{\small\begin{align*}
 G_{_N}(K,\mu_1,\varepsilon) &\geq \frac{I_0}{K}\left[\left(1-K|\mu_1|\right)^N + \left(1+K|\mu_1|\right)^N - 2\right]>0
\end{align*}}\\[-5pt]
with the last inequality following from the single-stock result; see Section I and the Appendix. \\[5pt]
\end{samepage}
{\it Case 2 (${\mu_1>0}$,~$N$~odd):} Assuming satisfaction of the theorem requirements~${G_{_N}(K,\mu_{min},\varepsilon_{max})>0}$ and~$G_{_N}(K,\mu_{max},\varepsilon_{max})>0$, and noting that~$\partial G_{_N}/\partial \varepsilon<0$, for~$\varepsilon\leq\varepsilon_{max}$, we obtain~$G_{_N}(K,\mu_{min},\varepsilon)>0$ and~$G_{_N}(K,\mu_{max},\varepsilon)>0$. Using this fact in conjunction with the definition of the critical uncertainty bound~$\varepsilon_c(\theta)$, we~obtain~$
\varepsilon_{max}< \min\{\varepsilon_c\left(K\mu_{min}\right),\varepsilon_c\left(K\mu_{max}\right)\}.$\\[7pt]
Invoking Lemma~$2$, we see that~$\varepsilon_c\left(\theta\right)$ is increasing when~$0<\theta< 2^{1/N}-1$ and from Lemma~$3$,~$\varepsilon_c\left(\theta\right)$ monotonically decreases after achieving a unique maximum at some~$\theta>2^{1/N}-1$. Thus, irrespective of the position of this maximal point, considering~$\theta=K\mu_{min}$ and~$\theta=K\mu_{max}$, we~have~$
\varepsilon_c\left(K\mu_1\right)\geq \min\{\varepsilon_c\left(K\mu_{min}\right),\varepsilon_c\left(K\mu_{max}\right)\}
$
for all~$\mu_{min}\leq \mu_1\leq \mu_{max}$.
Thus, for the arbitrarily chosen~$\mu_1$, it follows that~${\varepsilon_{max}<\varepsilon_c\left(K\mu_1\right)}$, which implies that~$G_{_N}(K,\mu_1,\varepsilon)>0$.\\[10pt]
{\it Case 3 (${\mu_1>0, N}$ even):} The first subcase which we consider is~when
$$
K>\frac{(2^{1/N}-1)}{\mu_{min}}
$$ holds. To show that~$G_{_N}(K,\mu_1,\varepsilon)>0$, we first note that the above strict inequality and the positivity of~$\mu_1$ and~$N$ being even implies that 
$$
{G_{_N}(K,\mu_1,\varepsilon)\geq \frac{I_0}{K}\left[\left(1+K\mu_1\right)^N-2\right]}>0.
$$ 
For the second subcase with~$K\leq\frac{2^{1/N}-1}{\mu_{min}}$,
{\small\begin{align*}
K&\leq \frac{1}{\mu_{min}}(1+\varepsilon_{max}) 
\end{align*}
}\\[-5pt]and~$G_{_N}(K,\mu_{min},\varepsilon_{max})>0$, it follows that
{\small\begin{align*}
 2-(1+K\mu_{min})^N\geq 0;\;\;
 1-K\mu_{min}(1+\varepsilon)\geq 0
 \end{align*}
}\\[-5pt]and
{\small\begin{align*}
 2-(1+K\mu_{min})^N&<(1-K\mu_{min}(1+\varepsilon))^N.
\end{align*}
}\\[-10pt]Hence,
$$
  \left[2-\left(1+K\mu_{min}\right)^N\right]^\frac{1}{N}<1-K\mu_{min}\left(1+\varepsilon_{max}\right),
$$
which, upon rearrangement and use of Lemma 1 leads~to
{\small\begin{align*}
 \varepsilon_{max}&<\frac{1-\left(2-\left(1+K\mu_{min}\right)^N\right)^\frac{1}{N}}{K\mu_{min}}-1
 =\varepsilon_c\left(K\mu_{min}\right).
\end{align*}
}\\[-5pt]Now, in the sub-subcase where~$\mu_1>(2^{1/N}-1)/K$, we have
{\small\begin{align*}
G_{_N}(K,\mu_1,\varepsilon)&\geq\frac{I_0}{K}\left[\left(1+K\mu_1\right)^N-2\right]>0. 
\end{align*}
}\\[-5pt]In the other sub-subcase,~$\mu_{min}\leq\mu_1\leq (2^{1/N}-1)/K$, from Lemma 2, we know that~$
\varepsilon_c\left(K\mu_{min}\right)\leq \varepsilon_c\left(K\mu_1\right).
$
Therefore,~${\varepsilon_{max}<\varepsilon_c\left(K\mu_1\right),}
$
implying that for the pair~$(\mu_1,\varepsilon)$, we have~$G_{_N}(K,\mu_1,\varepsilon)>0$. $\qed$
\subsubsection*{Proof of Corollary}
\input{Convergence_Simple}

%% file: necessity_noepsc.tex
To prove necessity, we assume~$G_{_N}(K,\mu_1,\varepsilon)>0$ for all admissible~$\mu_1$ and~$\varepsilon$, and consider two cases. For the case when~$N$ is odd, the claimed necessary condition
\begin{align*}
 G_{_N}(K,\mu_{min},\varepsilon_{max})>0;\;\;G_{_N}(K,\mu_{max},\varepsilon_{max})>0
\end{align*}
follows trivially from the fact that~$(\mu_{min},\varepsilon_{max})$ and~$(\mu_{max},\varepsilon_{max})$ are both admissible pairs.
For the case when~$N$ is even, the second necessary condition~${G_{_N}(K,\mu_{min},\varepsilon_{max})>0}$ is immediate using the same argument as for~$N$ odd. To complete the proof of necessity, we assume~${K\leq (2^{1/N}-1})/\mu_{min}$ and must show~that 
$$
{K\leq \frac{1}{\mu_{min}}(1+\varepsilon_{max})}.
$$
Indeed, proceeding by contradiction, if
$$
K>\frac{1}{\mu_{min}(1+\varepsilon_{max})},
$$
it is straightforward to verify that
{\small
$$
G_{_N}\left(K,\mu_{min},\frac{1}{K\mu_{min}}-1\right)= \frac{I_0}{K}\left[(1+K\mu_{min})^N -2\right]\leq 0,
$$}\\[-5pt]
which contradicts the assumed positivity of~$G_{_N}(K,\mu_1,\varepsilon)$.\\[8pt]

%% file: Convergence_Simple.tex
Given~${K>0}$, it suffices to show that for~$\varepsilon_{max}$ sufficiently small, the requirements of the RPE theorem are satisfied. 
For~$N$ odd, this follows since~$G_{_N}(K,\mu_{min},\varepsilon_{max})$ and~$G_{_N}(K,\mu_{max},\varepsilon_{max})$ are continuous functions of~$\varepsilon_{max}$,~$G_{_N}(K,\mu_{min},0)>0$ and~$G_{_N}(K,\mu_{max},0)>0$. That is, there exists a~$\delta_o>0$ such that for~$\varepsilon_{max}<\delta_o$, 
$$
G_{_N}(K,\mu_{min},\varepsilon_{max})>0;\;\;G_{_N}(K,\mu_{max},\varepsilon_{max})>0.
$$
For $N$ even and~$\varepsilon_{max}$ suitably small, the condition
$$
2^{1/N}-1\leq \frac{1}{1+\varepsilon_{max}}
$$
is easily seen to be satisfied. Now, arguing as in the case of $N$ odd, for $\varepsilon_{max}$ suitably small, we again~obtain
$$
G_{_N}(K,\mu_{min},\varepsilon_{max})>0.
$$
Thus, there exists~$\delta_e>0$ such that for~$\varepsilon_{max}<\delta_e$, the sufficient conditions for $N$ even are satisfied.~$\qed$